\documentclass[traditabstract]{aa}  
\usepackage{txfonts}
\usepackage{caption}
\usepackage{epstopdf}
\usepackage{tabu}
\usepackage{array,booktabs}
\usepackage[colorlinks=true,linkcolor=bluea,allcolors=blue]{hyperref}
\usepackage{siunitx}
\usepackage[switch]{lineno}
\usepackage{graphicx,longtable,lscape,natbib,amssymb,amssymb,amsmath}
\usepackage{float}
\bibpunct{(}{)}{;}{a}{}{,} % to follow the A&A style
\newcommand{\ltsima} {$\; \buildrel < \over \sim \;$}  
\newcommand{\gtsima} {$\; \buildrel > \over \sim \;$}  
\newcommand{\lta} {\lower.5ex\hbox{\ltsima}}  
\newcommand{\gta} {\lower.5ex\hbox{\gtsima}}

\newcommand{\WHz}{\>{\rm W}\,{\rm Hz}^{-1}}
\newcommand{\ergs}{\>{\rm erg}\,{\rm s}^{-1}}

\newcommand{\ergscm}{\>{\rm erg}\,{\rm s}^{-1}\,{\rm cm}^{-2}}
\newcommand{\ergscmA}{\>{\rm erg}\,{\rm s}^{-1}\,{\rm cm}^{-2}\,{\rm \AA}^{-1}}
\newcommand{\kms}{$\rm{\,km \,s}^{-1}$}
\newcommand{\lya}{Ly$\alpha\,$}

\usepackage{color}
\usepackage{xltabular}
\usepackage{longtable}

\usepackage{lineno}
\makeatletter
\renewcommand*\aa@pageof{, page \thepage{} of \pageref*{LastPage}}
\makeatother

\begin{document}

\title{The quest for high-redshift radio galaxies}
\subtitle{II. Discovery of the most powerful known radio galaxy at z=4.946.} 

\author{Barbara Balmaverde \inst{1}, Alessandro Capetti\inst{1}, Marco Chiaberge\inst{2,3}, Francesco Massaro\inst{4},  Miguel Coloma Puga\inst{1,4,5}, Ana Jimenez-Gallardo\inst{5}}
\institute{
  INAF - Osservatorio
  Astrofisico di Torino, Via Osservatorio 20, I-10025 Pino Torinese,
  Italy
\and
  Space Telescope Science Institute for the European Space Agency (ESA), ESA Office, 3700 San Martin Drive, 21218 Baltimore, USA
  \and
  Department of Physics \& Astronomy, Johns Hopkins University, Baltimore, USA.
  \and
  Dipartimento di Fisica, Universit\`a degli Studi di
  Torino, Via Pietro Giuria 1, 10125 (Torino), Italy
  \and European Southern Observatory, Alonso de Córdova 3107, Vitacura, Región Metropolitana, Chile}

  \abstract{As part of our quest for high-redshift radio galaxies (HzRGs), we present the results obtained from optical spectroscopy leading to the confirmation of one HzRG candidate selected using the Lyman break  technique. The optical emission, associated with the radio source TXS~2354+015, exhibits a prominent drop in the $r$ band, which is characteristic of sources at $z\sim5$. The optical spectrum shows a bright asymmetric emission line, with a large equivalent width of $\sim 900$\AA. Its identification with the \lya\ line is confirmed by the detection of the NV$\lambda1640$, leading to a redshift estimate of $z=4.946$. 
  Its broadband radio spectrum is well reproduced using a power law with a slope of 0.94. The radio luminosity at the rest frame frequency of 500 MHz is $6.2 \times 10^{29} \WHz$. This result makes this source the most powerful radio galaxy currently known.}

  \titlerunning{The quest for high-z radio galaxies.} \authorrunning{Balmaverde et al.}
\maketitle
  
  \section{Introduction}
\label{intro}

Powerful radio-loud active galactic nuclei (RLAGNs) represent the most extreme manifestation of accretion onto a supermassive black hole and play a crucial role in galaxy evolution. The energy input from the relativistic jets of RLAGNs can influence the star formation history in
their host and the energy balance of the intra-cluster medium (ICM;
e.g., \citealt{voit15,fabian12}). This active galactic nucleus (AGN) feedback is required in
numerical simulations to match their predictions with the observations of the galaxies' luminosity functions \citep{croton06}.
Radio-loud active galactic nuclei at high redshifts provide unique diagnostics about the conditions in the early Universe. For example, they represent beacons for finding
distant massive galaxies and proto-clusters, enabling us to explore
their properties and space density evolution (e.g., \citealt{wylezalek13,hatch14,magliocchetti22}). 

It is becoming increasingly clear that most high-z RLAGNs are obscured. \citet{volonteri11} first noted that the number of radio-loud (RL)
quasars is smaller than expected based on the number of blazars found at high redshifts. 
\citet{ghisellini16}  indicates that -- with respect to the number of blazars whose jets point toward us -- the number of expected parents (with jets pointing in other directions) is much lower than observed, based on the counts of
radio-detected sources in the FIRST+SDSS survey.
To solve this tension, they propose that a large fraction of high-redshift RLAGNs may be obscured by a bubble of dusty gas.
The optical emission -- including broad lines -- is absorbed, and
only the jet can pierce through this material. This scenario has been recently supported by \citet{capetti24}, who estimate that as much as 90\% of RLAGNs at z$>$3.5 are obscured in the optical and UV bands. Several studies (e.g., \citealt{merloni14,vijarnwannaluk22}) suggest that most high-redshift AGNs -- not just the RL subclass -- are obscured. Studying the cosmic X-ray background,
\citet{gilli07}, in particular, concluded that a population of heavily obscured Compton-thick AGNs is required to fit the X-ray data.
These results imply that the majority of the high-z RLAGNs appear as high-redshift radio galaxies (HzRGs), in which the active nucleus is hidden by circumnuclear absorbing material, rather than as radio quasars. 

Clearly, HzRGs are more difficult to detect than their unobscured quasars counterparts. 
Indeed, our
knowledge of HzRGs is extremely limited: only 37 radio galaxies (RGs) at $z>3$ are
listed by \citet{miley08} in their review (and just two at z$>$ 4.5). Only four spectroscopically confirmed HzRGs were subsequently added to this list \citep{jarvis09,yamashita20,saxena19}.\footnote{The redshift of the radio galaxy, TGSSJ1530+1049, at z=5.72 \citep{saxena18} has been recently revised to z=4.0 \citep{saxena25}.}

In addition, these HzRGs have been selected using
different methods; thus, they do not form a homogeneous sample. Most HzRGs have been identified by obtaining optical spectroscopy of ultra-steep spectrum (USS) radio sources, defined as objects with a radio spectral index $\alpha$ 
\footnote{The spectral index, $\alpha$, is defined as a radio spectrum in the form, $F_\nu \propto \nu^{-\alpha}$.} 
greater than 1.3, following the definition of \citet{saxena18}, or the looser criterion ($\alpha > 1$) proposed by \citet{rottgering97}. However, identifications of HzRGs based on the Lyman break technique yield sources with flatter radio spectra that do not conform with the USS criterion \citep{jarvis09,yamashita20,capetti25}. This suggests that the USS technique selects only a subpopulation of HzRGs. This conclusion is supported by the analysis by \citet{ker12}, who concluded that a steep-spectrum selection excludes more than half of the HzRGs.

\begin{figure*}
   \centering
  \includegraphics[width=0.97\textwidth]{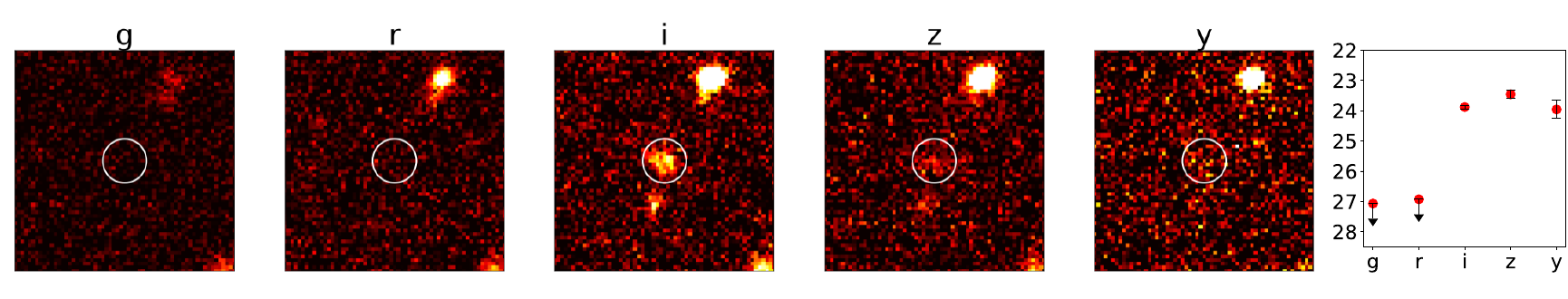}
    \caption{Optical cutouts in the five Subaru HSC SSP survey bands. The fields of view are 10\arcsec$\times$10\arcsec. The white circle marks the position of the candidate HzRG. Right plot: AB magnitudes in the five HSC-SSP bands. The arrows in the $g$ and $r$ bands indicate upper limits.}
     \label{cutouts}
\end{figure*}

\begin{figure*}
   \centering
  \includegraphics[width=0.47\textwidth]{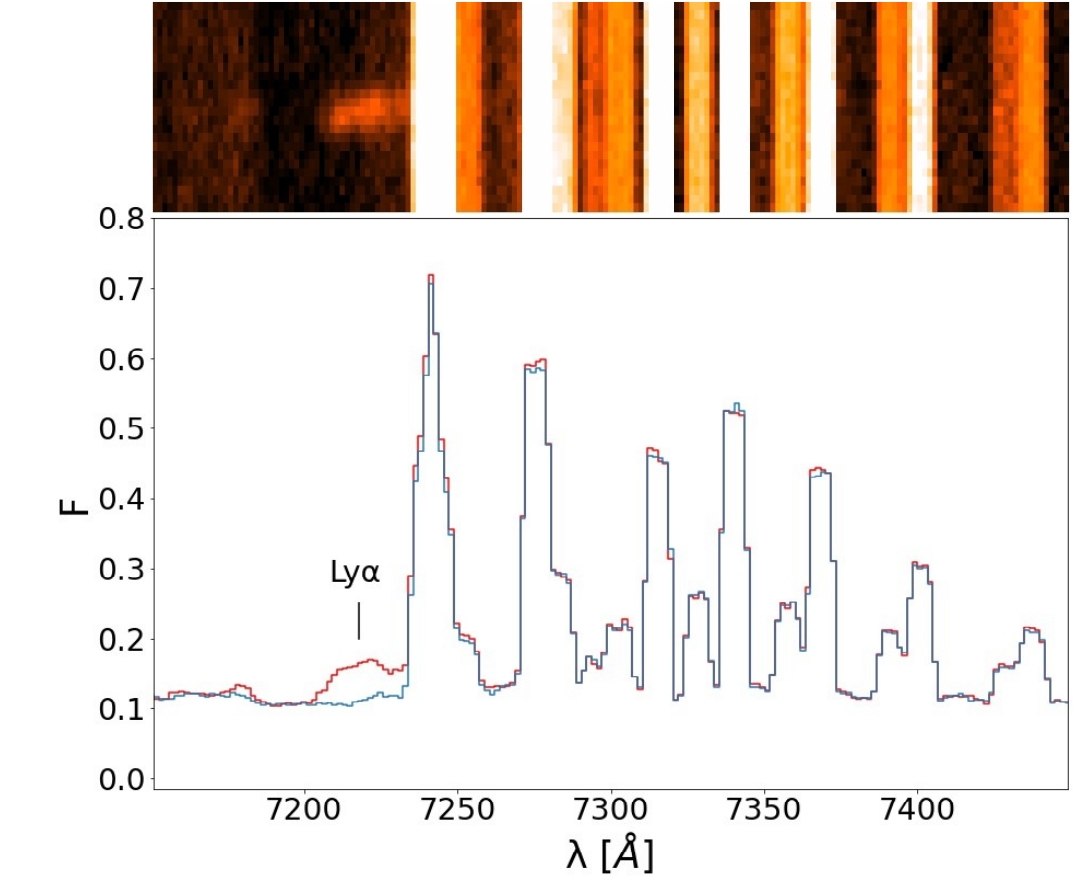}
  \includegraphics[width=0.52\textwidth]{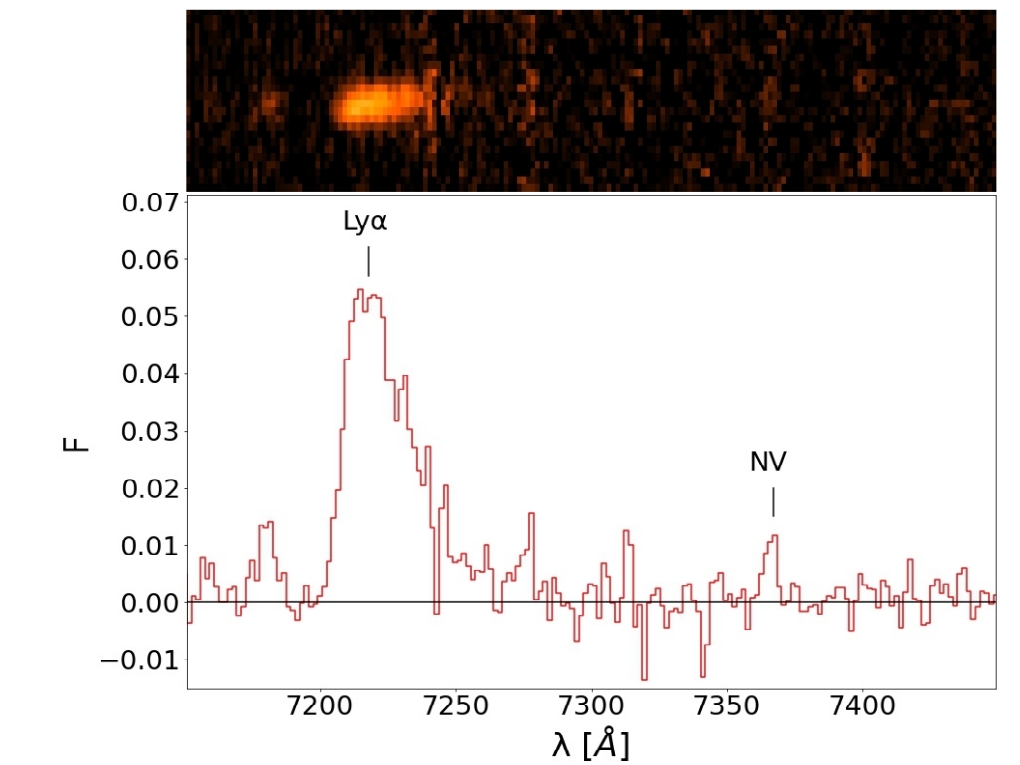}
    \caption{Top left: 6\arcsec\ wide 2D spectrum in the 7150-7450 \AA\ spectral range. Bottom: Spectrum extracted from a synthetic aperture of 5 pixels, with the location of the \lya\ and of the NV$\lambda1240$ emission lines marked. The flux scale is expressed in units of 10$^{-16} \ergscmA$. Right: Same as in the left panel following subtraction of the sky emission. }
     \label{vlt}
\end{figure*}

Several questions about these powerful RLAGNs thus remain unanswered. For example, while their co-moving space density increases dramatically by a factor of 100 - 1000 from the local Universe to $z \sim$ 2 \citep{willott01}, large uncertainties in their evolution at higher redshifts exist (\citealt{massardi10}). Another issue that can be addressed by studying HzRGs is the properties of their host galaxies. For example, we can test whether, as observed in nearby RLAGNs, these sources represent the high-luminosity end of the galaxies' at high redshifts, as suggested by   \citet{rocca04}. This test cannot be performed on RL quasars due to the strong contamination caused by nuclear light.

Our aim is to build a statistically sound sample of z $>$ 3 radio galaxies that would enable us to robustly explore their properties. In \citet{capetti25}
we presented the results of a pilot program aimed at obtaining spectroscopic confirmation of HzRGs candidates selected based on the Lyman break technique (e.g., \citealt{steidel95}). We observed $g$-dropout sources, expected to lie at 3.0 $<z<$4.5, and discovered four HzRGs with $3.3 < z < 3.8$. 

We here present the first results of our quest for sources at even higher redshifts, based on observations of $r$-dropouts, expected to lie at $4.5 < z < 5.3$.
We focus on the results obtained from optical spectroscopy of one such candidate associated with the radio source TXS~2354+015. The detection of a prominent emission line, readily identified with the \lya line, led us to locate it at a redshift $z=4.946$. 

This paper is organized as follows. In Sect. \ref{selection} we outline our selection methods for the HzRG candidates and present the main properties of the optical source associated with TXS~2354+015. The VLT observations and results obtained are presented in Sect. \ref{data}. The available radio data are presented in Sect. \ref{radio}.  Our results are discussed in Sect. \ref{discussion}.
We adopted the following set of cosmological parameters: H$_{0}=69.7$ km s$^{-1}$ Mpc$^{-1}$ and $\Omega_{m}=0.286$ (\citealt{bennett14}).  
At the redshift of 
TXS~2354+015, these values yield a scale of 6.45 kpc arcsec$^{-1}$.

\section{Target selection}
\label{selection}

The HzRGs candidates were selected using
large-area radio and optical surveys. On the optical side, we
focused on the data from the Hyper Suprime-Cam Subaru Strategic Program survey (HSC-SSP; \citealt{aihara18}), which covers 1400 deg$^2$. The catalog depth at 5$\sigma$ is $g$ = 26.5, $r$ = 26.1, $i$ = 25.9, z = 25.1, and Y = 24.4 mag. It is well suited
to search for HzRGs.

On the radio side, we used data from the Tata Institute of Fundamental
Research Giant Metrewave Radio Telescope (GMRT; \citealt{swarup91,intema17}) TGSS. The TGSS was conducted at 150 MHz, with a spatial resolution of $\sim
25\arcsec$, a flux density limit of 17.5 mJy, and covers the region $\delta > -53^\circ$. It detected 623,604 sources. Our choice to select candidates at low frequencies is motivated by two factors: the need to sample the rest-frame gigahertz radio emission in high-redshift sources, and the suggestion that HzRGs are mostly found among USS
sources, which are most detectable at low frequencies.

We considered the 20,642 TGSS sources within the HSC-SSP wide-field footprint, limiting the selection to those with a flux density F $>30$ mJy. The spatial resolution of the TGSS images is insufficient to accurately locate and identity the host galaxy of the radio source. We thus used the catalog of radio sources from the Karl G. Jansky Very Large Array Sky Survey (VLASS, \citealt{lacy20}). The VLASS includes multi-epoch images of the sky at 3 GHz with $\delta>-40^\circ$, achieving a resolution of 2\farcs5 and a
rms of 70 $\mu$Jy/beam in the co-added data. We extracted all sources from the VLASS catalog within a distance of 15\arcsec\ from the TGSS positions. In cases where multiple VLASS components were associated with a given TGSS source, we included all of them in the list of radio positions. 

The search for the HSC-SSP optical counterparts was conducted using
a matching radius of 1\farcs5 around each VLASS radio components, a value motivated by the relative astrometric uncertainties of these radio and optical surveys. 
From this list we selected r-dropout sources based on the criteria from \citet{ono18} and \citet{pouliasis22}. Measurements of the $r-i$ versus $i-z$ colors requires detecting sources in the $i$ band. Objects with upper limits in either the $r$ or the $z$ band may be selected as r-dropouts, provided that the corresponding color limits fall within the appropriate region. Among the r-dropouts found, we here focus on the optical source associated with the radio source, TGSS~J235707.4+010543.

In Fig. \ref{cutouts} we present the optical cutouts centered on this source in the five bands obtained from the Subaru HSC-SSP survey. We also plot its composite model (cmodel) AB magnitudes, as also reported in Table \ref{table1}.\footnote {Cmodel magnitudes measure the total flux of sources by fitting a linear combination of exponential and de Vaucouleurs profiles.} 
The host of TXS~2354+015 in the $i$-band image is diffuse and slightly resolved (see Fig. \ref{cutouts}). A 2D Gaussian fit yields a Full Width at Half Maximum (FWHM) of 0\farcs9 $\pm 0\farcs2$, to be compared with that of nearby stars, which is 0\farcs48 $\pm 0\farcs07$. 

\begin{table}
\centering
%\caption{Psf-model and c-model optical and near infrared magnitudes.}
\caption{cmodel optical and near infrared magnitudes.}
\small
\setlength{\tabcolsep}{4pt}
\begin{tabular}{c c c c c c }
\midrule
 g & r & i & z & Y & W1\\
%$<$27.07 & $<$26.93 & 24.87 $\pm$ 0.01 & 24.71$\pm$ 0.01 & 24.81$\pm$ 0.01\\
 \multicolumn{5}{c}{[AB mags]} & [V mags] \\
\midrule  
$<$27.07 & $<$26.93 & 23.88$\pm$0.05 & 23.46$\pm$0.11 & 23.96$\pm$0.24 & 17.3$\pm$ 0.2 \\
\bottomrule
\label{table1}
\end{tabular}
\end{table}

\begin{table*}
\centering
\caption{TXS 2354+015: Main results.}
\label{table2}
\small
\setlength{\tabcolsep}{4pt}
\begin{tabular}{ l c c c c  c c}
\midrule
RA, DEC (J2000) & z &F(\lya) & L(\lya) & FWHM(\lya) & EW(\lya) & F(N~V)\\

%$<$27.07 & $<$26.93 & 24.87 $\pm$ 0.01 & 24.71$\pm$ 0.01 & 24.81$\pm$ 0.01\\
  &   &   [erg cm$^{-2}$ s$^{-1}$] & [erg s$^{-1}$] & [km s$^{-1}$] & $\AA$ & [erg cm$^{-2}$ s$^{-1}$]\\
\midrule  
359.2810, 1.8458 & 4.946 $\pm 0.001$ &  (0.91$\pm$ 0.03)$\times 10^{-16}$ & (3.0$\pm$0.1)$\times 10^{44}$ & 
1080$\pm$80  & 900$\pm$250 & (0.40$\pm$0.09)$\times 10^{-17}$ \\
\bottomrule
\end{tabular}
\end{table*}

\section{Observations and data analysis}
\label{data}

The VLT observations were conducted as part of program 116.28LA.001
using the FORS2 spectrograph on September 18, 2025. The seeing was 0\farcs60, 
and a 1\farcs3 wide slit was used. The observations employed the  
GRIS600RI grism, which covers the spectral 
range 5120 - 8450 \AA, in combination with the GG435 blocking filter. This setup provided a resolution of $\sim 300$ \kms\ at 7,000 \AA.
Two observations were obtained with an exposure time of 1218 s each. The target was acquired by pointing the telescope on a pivot source located 4\farcs4 NW of the target (see Fig. \ref{cutouts}) and orienting the slit so that it includes the target emission.
The long slit spectra were reduced using the standard Reflex pipeline \citep{freudling13}. The two reduced spectra were combined to remove cosmic ray events.  

The 2D spectrum and the spectrum extracted from a synthetic aperture of 5 pixels (corresponding to 1\farcs26) are shown in the left panel of Fig. \ref{vlt}. An excess of emission at the target's position, relative to the sky, is observed at $\lambda \sim 7220 \AA$. The sky emission was estimated by averaging its emission in a 3\farcs5\ wide region located 5\arcsec\ south of the target. The sky-subtracted spectrum is presented in the right panel of Fig. \ref{vlt}. A prominent emission line is seen. It is highly asymmetric, showing an extended red wing and a sharp drop on the blue side. Its FWHM is 1,080 $\pm$ 80 km s$^{-1}$, and its flux is
0.91$\pm0.03 \times 10^{-16} \ergscm$. The continuum on the red side of the line is barely detected, with a flux of $1.0\pm0.3\times 10^{-19} \, \ergscmA$. The line equivalent width relative to the red continuum is EW=900$\pm$ 250 \AA\ (see Table \ref{table2} for a summary of the TXS~2354+015 properties).

All these characteristics are typical of the Ly$\alpha$ emission line (see, e.g., \citealt{saxena18}). Further confirmation of this identification comes from the presence of a second emission line that we identify with the blue component of the NV$\lambda1240$ doublet. This line is expected to be, in the optical thin case, two times brighter than the red component and is detected in the VLT spectrum at 7367.6 \AA\ and with a flux of  0.040 $\pm 0.009 \times 10^{-16} \ergscm$.
The corresponding redshift based on the NV$\lambda1640$\ detection is z=4.946 $\pm$ 0.001, assuming an air wavelength for the 1239.16 \AA  \, line. The expected wavelength of \lya\ at this redshift is 7229 \AA, consistent with that observed. This makes TXS 2354+015 the second most distant radio galaxy after TN J0924-2201 at z=5.19 \citep{vanBreugel99}.

However, the putative NV line is superposed onto a bright sky line, and we therefore more accurately tested its reliability. We extracted a spectrum over the same spectral region but varying the position in the sky. Out of the 244 extracted spectra found, none exhibits a flux excess at a similar level. This analysis supports the correctness of the identification of the NV line.

\begin{figure}
   \centering
  \includegraphics[width=0.48\textwidth]{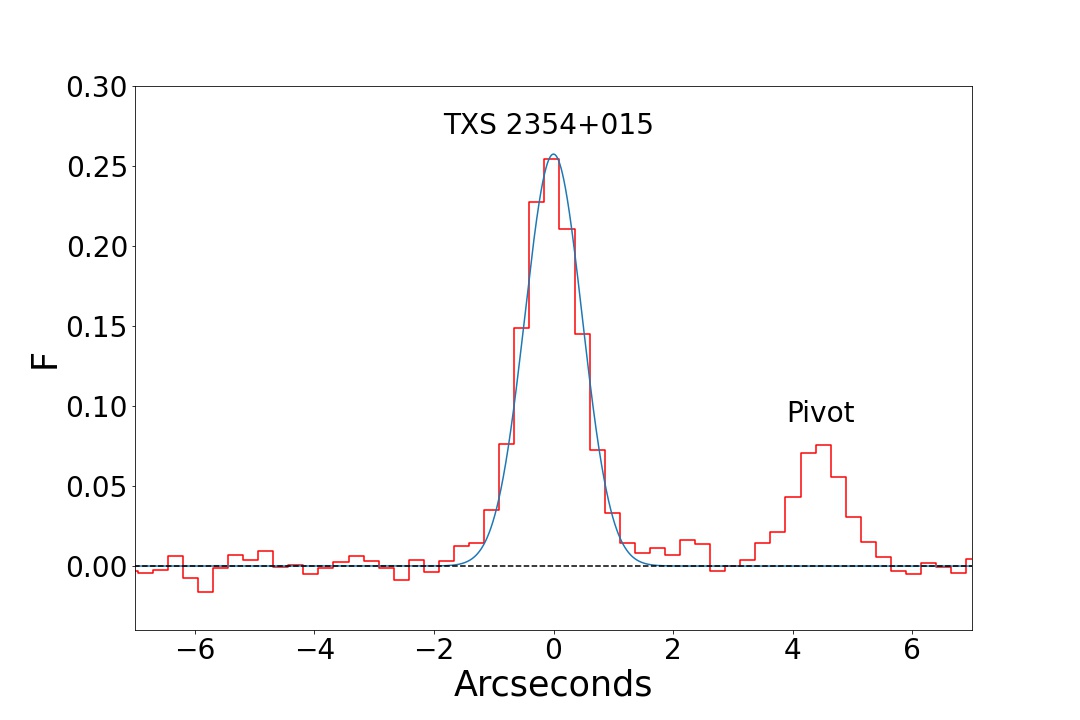}
    \caption{Spatial distribution of the emission extracted in the 7200-7250 \AA wavelength range. The main peak is produced by the \lya\ emission of TXS~~2354+015. The secondary peak arises from the pivot source used to acquire the target. The blue curve represents a Gaussian fit to the target's emission with a FWHM = 1\farcs13. The flux scale is expressed in units of 10$^{-16} \ergscm$.}
     \label{lyaext}
\end{figure}

In the spectrum the \lya\ emission is spatially extended (see Fig. \ref{lyaext}). In fact, its spatial profile is well reproduced by a Gaussian with a FWHM = 1\farcs13, while the seeing of the observations was 0\farcs60.

\section{Radio properties}
\label{radio}
\begin{figure*}
   \centering
\includegraphics[width=0.60\textwidth]{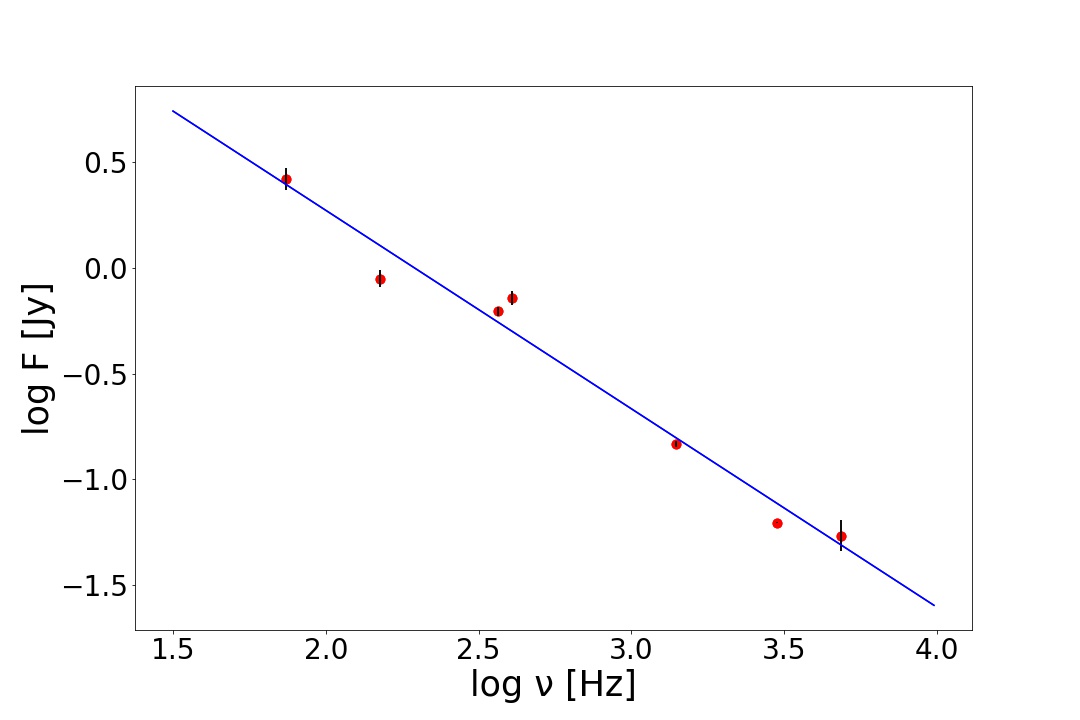} \includegraphics[width=0.37\textwidth]{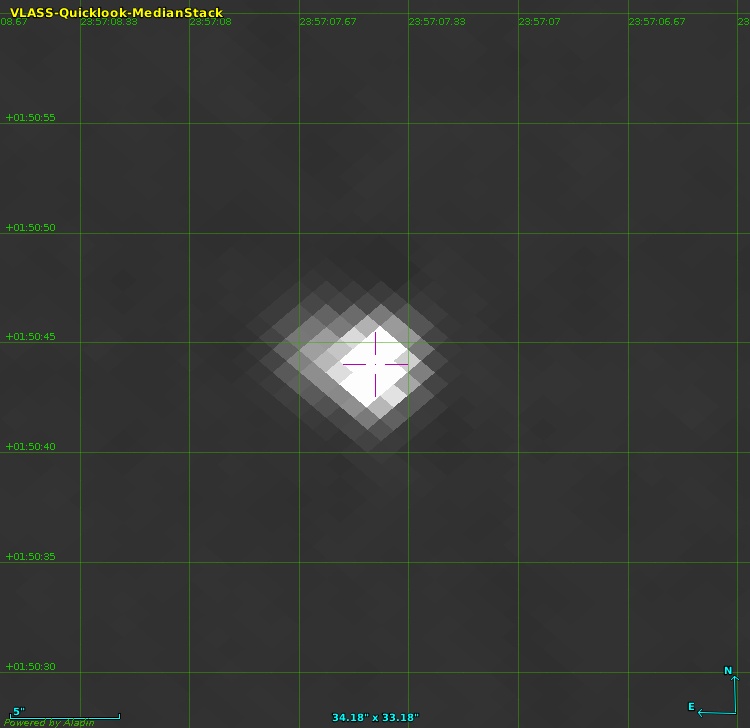}]
    \caption{Left: Radio spectrum of TXS~~2354+015 with data measurements from 74 MHz to 4.8 GHz. The solid line represents the best power-law fit to the data. Right: VLASS image of TXS~2354+015. The red cross marks the position of the optical source. The field of view is $\sim 30^{\prime\prime} \times 30^{\prime\prime}$.}
     \label{radiospec}
\end{figure*}

We collected radio flux density measurements for TXS~2354+015 from the NASA/IPAC Extragalactic Database, to which we added data from the TGSS and VLASS catalogs. The data span from 74 MHz to 4.8 GHz and are reported in Table \ref{table3}. The resulting radio spectrum is shown in the left panel of Fig. \ref{radiospec}.  
A single power law with an index $\alpha = 0.94\pm$0.03 accurately reproduces the spectrum, with no indications of a high-energy cutoff or a a low turnover frequency.

TXS~2354+015 is unresolved in all available observations. It is marginally resolved only in the
VLASS images, which have a 2\farcs5 resolution (corresponding to 16 kpc). An elliptical Gaussian fit yields a major axis of 3\farcs45$\pm 0\farcs01$ along a position angle of $71^\circ\pm1^\circ$ (see the right-hand image in Fig. \ref{radiospec}).\

\begin{table}
\centering
\caption{Radio flux density measurements.}
\label{table3}
\small
\setlength{\tabcolsep}{4pt}
\begin{tabular}{ c r }
\toprule
Frequency (MHz) & Flux density (mJy) \\
\midrule
74 &2620 $\pm$330\\ 
150 &888 $\pm$89  \\
365 &624 $\pm$31  \\
408 &720 $\pm$60 \\
1400 &146.8$\pm$ 4.4\\ 
3000 &62.4$\pm$ 0.4 \\
4800 &54 $\pm$10 \\

\bottomrule
\end{tabular}

\medskip
\end{table}

\section{Discussion}
\label{discussion}

Low-redshift powerful RGs exhibit a quasi-linear correlation between their luminosities of the low-frequency radio
emission and the narrow emission line (see, e.g. \citealt{baum89a,rawlings91}). This implies that the ratio of line to radio luminosity is effectively constant.
As reported by \citet{capetti25}, a similar behavior is seen in HzRGs,
which show a median ratio F$_{{\rm
    Ly}\alpha}/F_{150 {\rm MHz}} = 5\times 10^{-16}$ erg s$^{-1}$
cm$^{-2}$ Jy$^{-1}$ with a rms of $
\sim 3$. The estimated \lya\ flux for 
TXS~2354+015, based on its radio flux density, is $\sim 4 \times 10^{-16}$ erg s$^{-1}$, a factor of $\sim 5$ higher than the observed value. To establish whether this is due to saturation of the line emission at the highest radio luminosities, as observed at low redshifts \citep{capetti23}, a larger sample of $z \sim 5$ sources must be built.

The probability that TXS~2354+015 is a chance superposition of a high-redshift emission line galaxy and an unrelated radio source is  remote. The density of TGSS sources as bright as, or brighter than, TXS~2354+015 is only 0.07 per square degree. Moreover, the separation between the optical and VLASS positions is 0.2$^{\prime\prime}$, consistent with the relative astrometric uncertainties of the two surveys. Furthermore, the agreement between the observed and the expected \lya\ flux, given the radio flux density, further supports the correctness of the identification.

The \lya\ line is spatially extended, with a measured FWHM corresponding to $\sim$6 kpc. This result aligns with the known properties of the \lya\ nebulae (see, e.g., \citealt{wang23,coloma25}) that commonly have an extent exceeding tens of kiloparsec. In addition, the \lya\ emission is typically aligned with the radio axis. However, the observations of TXS~2354+015 were obtained with the slit oriented at the position angle PA=-30$^\circ$, almost perpendicular the radio PA (71$^\circ$). This suggests that the size measured by our long-slit observations underestimate its true extent. 
In the Subaru HSC-SSP survey, the host of TXS~2354+015 in the $i$-band image is slightly resolved with a FWHM of $\sim$5 kpc.
This value is significantly greater than
the typical radius of 1.5 kpc  galaxy at  z$\sim$4 (\citealt{bouwens04,ferguson04}). However, galaxies hosting radio AGNs are, at least in the nearby Universe, the most massive. Furthermore,
we cannot exclude that the measured size is driven by the presence of the extended \lya\ line, exhibiting an equivalent width EW similar to the filter width. 
Unfortunately, the source is too faint in the other bands, which are not contaminated by line emission, to measure its size and test this scenario. 

The radio spectrum of TXS~2354+015 is well reproduced with a single power law from 74 to 4800 MHz with a spectral slope $\alpha = 0.94$. Thus, it does
not conform to a USS classification. This result, combined with other HzRGs selected using the Lyman break technique \citep{jarvis09,yamashita20,capetti25}, indicates that an USS is not a general property of HzRGs. Although the preselection of USS radio sources boosts the fraction of high-z sources, this method unveils only a subsample of the HzRGs population.

TXS~2354+015 is marginally detected in the WISE W1 band with a VEGA magnitude W1=17.3 $\pm$ 0.2, corresponding to a luminosity of $0.9\times10^{46} \ergs$. The central wavelength of the W1 band is $\sim3.4  \, \mu$m with an $\sim 1  \, \mu$m width. At a redshift of 4.946, it includes several emission lines, the most prominent of them being [O~III]$\lambda5007$. We estimated its expected [O~III] luminosity by extrapolating the correlation between the radio and line emission. This was done by combining low-redshift radio galaxies \citep{buttiglione10} with HzRGs at 1.5$<$z$<$3.6 observed by \citet{nesvabda17}, yielding $L_{\rm[O~III]} \sim 1.5 \times10^{45} \ergs$.
Although this estimate was derived from an extrapolation at much higher radio power, it suggests significant contamination of the mid-infrared luminosity by emission lines. This implies that it cannot be used to, for example, derive a robust estimate of the mass of the host galaxy. Nonetheless, taking this estimate at face value and assuming $M/L = 0.1 M_\sun/L_\sun$ -- a value typical of a 100 Myr old stellar population \citep{bruzual03} -- we obtain a mass of $\sim 2 \times 10^{12} M_\sun$ for the host of TXS~2354+015. This value would place this galaxy at the very bright end of the local galaxy luminosity function (see, e.g., \citealt{cappellari13}). This estimate is clearly plagued by several large uncertainties, in particular by the age of the stellar population. For example, a dominant younger stellar population (10 Myr) would reduce the estimate by a factor of $\sim 10$. Improved coverage of the spectral energy distribution of this source is clearly required.

\citet{miley08} estimated the power at a rest-frame frequency of 500 MHz for all HzRGs known at that time, providing a robust reference for estimating their radio luminosity. The most powerful source is TN~J0924-2201, with a radio power $P_{500} = 3.2 \times 10^{29} \WHz$. The HzRGs discovered after the compilation of this 2008 review do not exceed this limit. We estimated the power at 500 MHz (84 MHz in the observer's frame) for TXS~2354+015 by interpolating the observed spectrum, finding $P_{500} = 6.2 \times 10^{29} \WHz$. TXS~2354+015, thus, appears to be the most powerful radio galaxy known to date.

\begin{acknowledgements}
We thank the anonymous referee for the useful comments that improved the paper. 
Based on observations made with ESO Telescopes at the La Silla Paranal Observatory under programme ID  116.28LA.001. The reduced spectra are available in the archive.
Based on data collected at the Subaru Telescope and retrieved from the HSC data archive system, which is operated by Subaru Telescope and Astronomy Data Center at National Astronomical Observatory of Japan. The Hyper Suprime-Cam (HSC) collaboration includes the astronomical communities of Japan and Taiwan, and Princeton University. 
The radio images were
retrieved from the NRAO VLA Archive Survey, (c) AUI/NRAO, available at http://archive.nrao.edu/nvas/. The National Radio Astronomy
Observatory is a facility of the National Science Foundation operated under cooperative agreement by Associated Universities, Inc.  BB acknowledges the prin INAF 2023 "The MURALES project: exploring AGN feedback in the most powerful radio loud active galactic nuclei".
Facilities: VLT:FORS2, VLA, Subaru Telescope.
\end{acknowledgements}

\bibliographystyle{aa} % style aa.bst
\bibliography{my} %

\end{document}